\documentclass[a4paper,11pt]{article}
\pdfoutput=1 

\usepackage{jheppub} 

\usepackage[T1]{fontenc} 
\usepackage{amsmath}
\usepackage{amsfonts}
\usepackage{amsmath}
\usepackage{amsxtra}
\usepackage{amstext}
\usepackage{amssymb}
\usepackage{slashed}
\usepackage{graphicx}
\usepackage{cleveref}
\usepackage{color}
\usepackage[utf8]{inputenc}

\numberwithin{equation}{section}

\newcommand{\beq}{\begin{equation}}
\newcommand{\eeq}{\end{equation}}
\newcommand{\bea}{\begin{eqnarray}}
\newcommand{\eea}{\end{eqnarray}}

\newcommand{\nn}{\nonumber}
\newcommand{\<}{\langle}
\renewcommand{\>}{\rangle}

\newcommand{\im}{\mathrm{Im}\,}
\newcommand{\re}{\mathrm{Re}\,}

\DeclareMathOperator{\Tr}{Tr}

\newcommand{\svev}{\langle S \rangle}

\title{\boldmath 

Smearing and Unsmearing KKLT AdS Vacua}

\author{Mariana Gra\~na,}
\author{Nicolas Kovensky,}
\author{Dimitrios Toulikas.}

\affiliation{Institut de Physique Th\'eorique, Universit\'e Paris Saclay, CEA, CNRS, Orme des Merisiers, 91191 Gif-sur-Yvette Cedex, France.}

\emailAdd{mariana.grana@ipht.fr}
\emailAdd{nicolas.kovensky@ipht.fr}
\emailAdd{dimitrios.toulikas@ipht.fr}

\abstract{Gaugino condensation on D-branes wrapping internal cycles gives a mechanism to stabilize the associated moduli. According to the effective field theory, this gives rise, when combined with fluxes, to supersymmetric AdS$_4$ solutions. In this paper we provide a ten-dimensional description of these vacua. We first find the supersymmetry equations for type II AdS$_4$ vacua with gaugino condensates on D-branes, in the framework of generalized complex geometry. We then solve them for type IIB compactifications with gaugino condensates on smeared D7-branes.  
We show that supersymmetry requires a (conformal) Calabi-Yau manifold and imaginary self-dual three-form fluxes with an additional (0,3) component. The latter is  proportional to the cosmological constant, whose magnitude is determined by the expectation value of the gaugino condensate and the stabilized volume of the cycle wrapped by the branes. This confirms, qualitatively and quantitatively, the results obtained using effective field theory. We find that exponential separation between the AdS and the KK scales seems possible as long as the three-form fluxes are such that their (0,3) component is exponentially suppressed.
As for the localized solution, it requires going beyond SU(3)-structure internal manifolds. Nevertheless, we show that the action can be evaluated on-shell without relying on the details of such complicated configuration. We find that no "perfect square" structure occurs, and the result is divergent. We compute the four-fermion contributions, including a counterterm, needed to cancel these divergencies.

}

\begin{document} 
\maketitle
\flushbottom

\newpage

\newpage
\section{Introduction}
\label{sec:intro}

Gaugino condensation on D-branes wrapping internal cycles of string compactifications provides a mechanism for stabilizing their associated moduli. Indeed, the gauge coupling, appearing in the superpotential, depends on the corresponding volumes. This is particularly handy in type IIB compactifications on  Calabi-Yau manifolds. The three-form fluxes threading the internal three-cycles are routinely included, and provide a potential for their sizes, parameterised by the complex structure moduli. On the other hand, K\"ahler moduli, which define the sizes of the even cycles, are unfixed at the perturbative level, hence the non-perturbative contributions coming from gaugino condensates on D-branes are crucial. For type IIB compactifications with three-form fluxes and O3/O7 planes, such as the ones considered in this paper, the supersymmetric branes are D7-branes wrapped on calibrated four-cycles $\Sigma_4$.

This proposal for supersymmetric compactifications with fully stabilized moduli 
was put forward by Kachru, Kallosh, Linde and Trivedi (KKLT) in  \cite{Kachru:2003aw}.  More explicitly, complex structure moduli are assumed to be fixed by fluxes\footnote{This mechanism was conjectured not to work for a large number of moduli because of the tadpole cancellation condition \cite{Bena:2020xrh}.}, and at a lower energy scale one uses effective field theory to study K\"ahler moduli. The corresponding F-term conditions then lead to a supersymmetric AdS$_4$ solution. As long as the (0,3) fluxes can be fine tuned to give a very small contribution to the superpotential, comparable to the non-perturbative one, the resulting cosmological constant is exponentially small, while the K\"ahler moduli are fixed at a large value. This is a prominent example of scale separation, which violates the Swampland conjecture formulated in \cite{Gautason:2018gln}.
Importantly, a family of explicit examples were constructed recently in \cite{Demirtas:2019sip,Demirtas:2020ffz,Demirtas:2021nlu,Demirtas:2021ote}.

In this paper we provide the ten-dimensional description of these supersymmetric AdS$_4$ vacua with fluxes and gaugino condensates captured by the EFT of \cite{Kachru:2003aw}. For that, we first derive the supersymmetry equations for AdS$_4$ compactifications in the presence of gaugino condensates, combining different elements that appeared in the literature  \cite{Grana:2005sn,Koerber:2008sx,Dymarsky:2010mf,Bena:2019mte,Kachru:2019dvo} and bringing them together into a consistent picture. We work within the framework of Generalized Complex Geometry (GCG) \cite{Grana:2005sn}. This is necessary since, as shown in \cite{Koerber:2007jb,Dymarsky:2010mf,Bena:2019mte}, the backreaction of the gaugino condensate breaks the SU(3) structure of Calabi-Yau manifolds down to a more general so-called "dynamic SU(2) structure", best understood in terms of generalized complex structures. Requiring ${\cal N}$=1 supersymmetry gives three equations, involving the generalizations of the complex structure and the complexified K\"ahler structure. The first two of these conditions were shown to be equivalent to the F-flatness conditions for the K\"ahler and complex structure moduli, while the third one corresponds to a D-flatness condition \cite{Koerber:2007xk} of the effective theory of compactifications on generalized geometries \cite{Grana:2005sn,Benmachiche:2006df}. In the presence of gaugino condensates, these equations get modified. While the modification of the first supersymmetry condition was understood in \cite{Koerber:2008sx,Dymarsky:2010mf} in terms of the backreaction on the geometry itself, that of the third equation, which involves the RR fluxes, is more subtle. This was considered first in \cite{Dymarsky:2010mf,Kachru:2019dvo} under some approximations. Here we take an alternative route and find the generalized geometry extension for gaugino condensation in any type II branes wrapping a calibrated cycle. We do so by building on the analysis of gaugino mass terms presented in \cite{Grana:2020hyu}. Furthermore, we argue for self-consistency of the whole set of equations, and consistency with the four-dimensional effective theory\footnote{At this level, our results are consistent with the discussion of \cite{Kachru:2019dvo}, see their Appendix A in the latest version.}.

We then solve these modified supersymmetry equations  
for type IIB AdS$_4$ compactifications with gaugino condensates on \textit{smeared} D7-branes. We find that the solution is surprisingly simple, and shares many features with its Minkowski counterpart without gauginos. More precisely, we find that the internal manifold is still  (conformal) Calabi-Yau, and three-form fluxes are still imaginary self-dual. Nevertheless, they contain an additional (0,3) piece, which turns out to be proportional to the non-vanishing cosmological constant,  whose value is in turn dictated by the vacuum expectation value (VEV) of the gaugino bilinear.
These features of the ten-dimensional solution reproduce the expectations from the four-dimensional effective field theory analysis in \cite{Kachru:2003aw} not only qualitatively but also quantitatively. We take this as strong evidence confirming both our modified supersymmetry equations and also the applicability of the EFT for  finding supersymmetric vacua.

We further analyze scale separation (see e.g. \cite{Apers:2022zjx} and references therein) in the smeared solution, finding two relevant scales, the dilaton and the volume. We show that exponential scale separation can be achieved as long as the gaugino vev is very small, which happens at weak coupling in the gauge theory, and as long as  one can cook up fluxes giving rise to an equally small (0,3) component. 

The equations for gaugino condensates on localised D7-branes are, not surprisingly, much harder to solve. We leave the study of the complete solution for future work, and only comment on some of the key features of the partial solutions obtained in \cite{Dymarsky:2010mf,Kachru:2019dvo}. Nevertheless, we consider in detail the on-shell value of the bulk plus brane action. Since some components of the flux, as well some derivatives of the pure spinors that encode the generalized complex geometry contain delta functions that localize them on the $\Sigma_4$ cycle, one might worry whether such on-shell action is divergent (i.e.~whether it has terms involving squares of delta functions). This question was raised recently and discussed in several papers \cite{Hamada:2018qef,Hamada:2019ack,Gautason:2019jwq,Kallosh:2019oxv,Hamada:2021ryq,Kachru:2019dvo}, without reaching a common conclusion. We show that it is possible to compute the on-shell action without knowing the details of the solution, assuming it solves the modified supersymmetry conditions. 
More precisely, we compute the on-shell action up to two-fermion terms using the results obtained in \cite{Grana:2020hyu} for the relevant terms of the D-brane action, as well as the expression for the bulk ten-dimensional supergravity Lagrangian for generalized geometry compactifications in terms of fluxes and derivatives of the pure spinors, obtained in \cite{Lust:2008zd}. We find that the action is indeed divergent, and compute the coefficients of the terms involving one and two delta functions. The former should be cancelled by four-fermion terms in the D-brane action, while the latter indicate that a counterterm must be included as well.

The paper is organized as follows. In section \ref{sec:susywogauginos}
we review the supersymmetry equations without gaugino condensates, their equivalence with supersymmetry conditions of the four-dimensional description, their self-consistency and the four-dimensional Minkowski 
solutions. In section \ref{sec:susywgauginos} we present the supersymmetry equations with gaugino condensates and the arguments that lead to them, including self-consistency of the equations. In section \ref{sec:smeared} we construct the solution for smeared branes, compare with the effective four-dimensional theory and discuss scale separation.
In section \ref{sec:localised} we discuss the main features of the localised solution, and the divergence of the on-shell action. Finally, we discuss our results in section \ref{sec:conclusions}.

\section{Supersymmetry conditions for type II AdS$_4$ vacua from 10D} 
\label{sec:susywogauginos}

In this section we review how type II superstring theory Mink$_4$ and AdS$_4$ vacua with classical sources (such as D-branes or O-planes) are described from the ten-dimensional point of view using the language of generalized complex geometry. We also establish the conventions that we will use throughout this paper, following \cite{Koerber:2007xk,Koerber:2010bx}.  

\subsection{Supersymmetric AdS$_4$ vacua without gaugino condensates}

The GCG conditions for type II string flux compactifications to four-dimensional flat and AdS vacua preserving ${\cal N}=1$ supersymmetry were found originally in \cite{Grana:2005sn}. They are written in terms of two polyforms, denoted $\Psi_\pm$, which characterize the internal geometry. For type IIB these conditions read 
\begin{subequations}
\label{susy}
 \begin{align}
d_H \left(e^{3A-\phi} \Psi_-\right) &= 2 i \mu e^{2A-\phi} \im \Psi_+  \label{susy1}\\
d_H \left(e^{2A-\phi} \im \Psi_+ \right) &= 0 \label{susy2}\\
d_H \left(e^{4A-\phi} \re \Psi_+ \right) &= 3 e^{3A-\phi} \re \left[ \bar{\mu} \, \Psi_- \right] + e^{4A} *_6 \alpha( F) \, , \label{susy3}
 \end{align}
\end{subequations}
while for type IIA one just has to interchange $\Psi_+$ with $\Psi_-$. In Eqs.~\eqref{susy} $\phi$ is the dilaton, while $A$ is the warp factor, such that, in the string frame, the 10D metric splits as follows: 
\beq \label{10Dmetric}
ds_{10}^2=e^{2A(y)} g_{\mu\nu}(x) dx^\mu dx^\nu + h_{mn}(y) dy^m dy^n \ ,
\eeq
where $g_{\mu \nu}$ describes the extended  AdS$_4$/Mink$_4$ directions. Moreover,  $\mu$  is related to the cosmological constant by
\beq
\Lambda=-3|\mu|^2 \, , \
\eeq
and 
\beq \label{dH}
d_H \equiv d + H\wedge \ 
\eeq
is the $H$-twisted exterior derivative.
The polyforms or pure spinors $\Psi_{\pm}$ are defined as 
\begin{equation} 
\label{purespinors}
    \Psi_{\pm} \equiv -\frac{8 i}{||\eta||^2} \sum_p \frac{1}{p!} \eta^{2\dagger}_{\pm} \gamma_{m_1 \dots m_p} \eta^1_+ \, dy^{m_1} \wedge \dots \wedge dy^{m_p} \ , 
\end{equation}
where $\eta^1$ and $\eta^2$ are two globally defined spinors on the internal manifold, that can become parallel at certain loci, and whose norm is related to the warp factor as $||\eta^1||^2 = ||\eta^2||^2 = e^{A}$.
By using properties of spinor bilinears in six dimensions it is easy to see that $\Psi_-$ and $\Psi_+$ are sums of odd and even $p$-forms, respectively, and satisfy the self-duality condition
\beq \label{*Psi}
*_6 \alpha(\Psi_\pm) = i \Psi_\pm   \,,  \quad  \alpha(\omega_q)=(-1)^{\frac{q(q-1)}{2}}\omega_q \,.
\eeq
Finally, the polyform $ F$ accounts for the RR fluxes on the internal manifold, which are related to those with external legs by self-duality. More explicitly, the total\footnote{Here we use the democratic formulation \cite{Bergshoeff:2001pv}, with the polyform notation, where $F = \sum F_q$ with  $q=1,3,5,7,9$ ($q=0,2,4,6,8,10$) for type IIB (IIA).} RR flux $\hat{F}$ is 
\beq
\label{Fintext}
\hat F = { F} + e^{4A} \mathrm{vol}_4 \wedge \tilde{F},
\eeq
such that 
\beq \label{sdF6}
\tilde F= *_6 \, \alpha({ F}) \ .
\eeq

As discussed in \cite{Grana:2020hyu}, the third supersymmetry condition, namely Eq.~\eqref{susy3}, can be understood in terms of the generalized flux 
\begin{equation}
    G \equiv F + i e^{-4A}d_H\left(e^{4A-\phi} \re \Psi_+\right)
    \label{defG}
\end{equation}
as follows:
\begin{equation}
    (1-i *_6 \alpha) G = 3 i e^{-A-\phi} \bar{\mu} \Psi_- \, . 
    \label{IASD0}
\end{equation}
Upon restricting to internal manifolds with a well-defined SU(3) structure and compactifications with $\mu=0$, this reduces to the usual imaginary-self-duality (ISD) condition on the three-form flux
\begin{equation}
    G_3 = F_3 + i e^{-\phi} H \, .
\end{equation} 
In this sense, any AdS$_4$ solution to the (classical) supersymmetry conditions \eqref{susy} must include some IASD contributions to the generalized flux $G$. 

A supersymmetric configuration is a solution to the equations of motion iff Eqs.~\eqref{susy} are satisfied and all fluxes satisfy the corresponding Bianchi identities 
\begin{equation}
dH=0\ , \quad    d_H F = d F + H \wedge F= \delta_{\rm{Dp}}\,,
\label{Bianchi}
\end{equation}
where the possible sources encoded in $\delta_{\rm{Dp}}$ are either D-branes or O-planes. Indeed, the EOM for the fluxes follow directly from the supersymmetry conditions. In the polyform language, they read 
\begin{equation}
\label{EOMF}
    d_H \left[e^{4A} *_6 \alpha (F)\right]=0 \,, 
\end{equation}
as can be seen by applying $d_H$ to Eq.~\eqref{susy3} and using that $d_H^2=0$, together with \eqref{susy1}. Although the EOM for $H$ is more cumbersome in the GCG language, it was shown in  \cite{Koerber:2007hd} that it follows from the hodge-dual of the three-form component of Eq.~\eqref{susy3}. We will come back to this later on.

 \subsection{Supersymmetry conditions from the 4D EFT and superpotential}
 \label{sec:susy4D}
 
The conditions in Eq.~\eqref{susy} are equivalent to requiring that the supersymmetry variations of the gravitino and the dilatino vanish. Importantly, it was shown in \cite{Grana:2005ny,Grana:2006hr,Benmachiche:2006df,Koerber:2007xk,Koerber:2008sx,BilalCassani}  that they can also be understood as D- and F-flatness conditions in the four-dimensional effective action for the scalars in vector and chiral multiplets. In type IIB, the former come from deformations of the complex structure, while the latter are combinations of the RR axions and the K\"ahler deformations. In the language of GCG, the complex structure is encoded in $\Psi_-$, while the K\"ahler moduli are contained in $\mathrm{Re} \, \Psi_+$. One therefore defines the holomorphic fields  
\beq \label{Z,T}
Z = e^B e^{3A-\phi}  \Psi_- \ , \qquad T=e^B ( C + i e^{-\phi} \, \mathrm{Re}\,  \Psi_+ ) \ , 
\eeq
where $C$ are the RR gauge potentials, i.e.~$d_H C =  F $. These are precisely the combinations whose exterior derivatives appear in Eqs.~\eqref{susy1} and  \eqref{susy3}, respectively. The argument of the derivative in \eqref{susy2} can be formally thought of  as a function of $Z$ and $T$ \cite{Benmachiche:2006df}. In order to build the low-energy effective action one would need to specify the (a priori massless) deformations of these geometric objects, which contain all relevant information about the internal metric, the warp factor $A$, the dilaton $\phi$ and the $B$-field. 
Here however what one does is to build a superpotential in terms of the full pure spinors, which involves an infinite number of fields, including all the Kaluza-Klein modes. In this sense, we can define a ten-dimensional superpotential, which is given by \cite{Koerber:2007jb,Grana:2005ny,Grana:2006hr,Benmachiche:2006df} 
\begin{equation}
    W_{\rm GCG} =  
    \pi  \int_{M_6} \<Z, d \, T\> = \pi \int_{M_6} \<e^{-B}Z, G\> \ ,    
    \label{W10D}
\end{equation}
where $G$ was defined in \eqref{defG}.
The brackets in Eq.~\eqref{W10D} correspond to the so-called Mukai pairing, 
\begin{equation} \label{Mukai}
    \<A,B\> \equiv \left[A \wedge \alpha(B) \right]_{6} \ ,
\end{equation}
where one only includes the 6-form component. As discussed below, the variations of $W_{\rm GCG}$ as function of $Z$ and $T$ then vanish iff Eqs.~\eqref{susy} are satisfied. 


\subsection{K\"ahler potential, cosmological constant, and  self-consistency}
\label{sec:Wvev}

Having defined the generalized superpotential in Eq.~\eqref{W10D}, we now turn to the K\"ahler potential. This can be understood in terms of
\begin{equation}
    {\cal{N}} = 4 \pi  \int_{M_6}  e^{2A-2\phi} \mathrm{vol}_6 \,,  
    \label{Ndef}
\end{equation}
which is the constant appearing in front of the Einstein-Hilbert term of the effective four-dimensional action, thus setting the corresponding Planck scale. This defines the K\"ahler potential in the (Einstein frame) 4D supergravity language \cite{Koerber:2007xk} 
 \begin{equation}
    {\cal{K}} = -3 \log {\cal{N}}.
\end{equation}
By using the pure spinor normalizations 
\begin{equation}
    \langle \Psi_+, \bar{\Psi}_+ \rangle
    = 
    \langle \Psi_-, \bar{\Psi}_- \rangle
    = - 8 i \mathrm{vol}_6 \,, 
    \label{purespinorVol}
\end{equation}
the K\"ahler potential can be expressed as 
\begin{equation}
    {\cal{K}} = -2 \log i \int_{M_6} e^{2A}
    \langle t, \bar{t}\rangle 
    - \log i \int_{M_6} e^{-4A}
    \langle z, \bar{z}\rangle - 3 \log \frac{\pi}{2} \,, 
\end{equation}
with $t = e^{-\phi} \Psi_+$ and $z = e^{3A-\phi} \Psi_-$. 
This allows one to interpret Eqs.~\eqref{susy1} and \eqref{susy3} as the F-flatness conditions associated to the variations of $W_{\rm GCG}$ with respect to $T$ and $Z$, respectively. The remaining condition \eqref{susy2} (which is automatically satisfied for compactifications with non-zero $\mu$) can similarly be interpreted as a D-term condition. 

For this interpretation to hold, and for Eqs.~\eqref{susy3} to give a self-consistent system of equations in terms of $W_{\rm GCG}$, the cosmological constant as denoted by $\mu$ must correspond to the on-shell value of the superpotential. More precisely, we should have 
\begin{equation}
    \langle W_{\rm GCG} \rangle = \mu \,  {\cal{N}}\,.
    \label{muWonsell}
\end{equation}
We now review how this is derived. Let us split the contributions to the on-shell superpotential \eqref{W10D} as follows: 
\begin{equation}
\langle W_{\rm GCG} \rangle = \pi
    \int_{M_6} \<e^{3A-\phi} \Psi_-,  F \> + \pi \int_{M_6} \<e^{3A-\phi} \Psi_-,  i d_H [ e^{-\phi} \mathrm{Re} \Psi_+]\>
    \, . 
    \label{Wonshell1}
\end{equation}
In order to relate the term proportional to $F$ with Eq.~\eqref{susy1} we use that the Mukai pairing satisfies $\langle \Psi,\Phi\rangle = \langle *_6 \alpha (\Psi), *_6 \alpha (\Phi)\rangle$ for generic polyforms $\Psi,\Phi$. Since, the pure spinors are ISD (see Eq.~\eqref{*Psi}), by using Eq.~\eqref{susy3} we have
\begin{equation}
    \<e^{3A-\phi} \Psi_-,  F \> = 
    i \<e^{3A-\phi} \Psi_-,  \tilde{F} \>
    = 
    i \<e^{3A-\phi} \Psi_-,  
    e^{-4A}d_H \left(e^{4A-\phi} \re \Psi_+ \right) - 3 e^{-A-\phi} \re \left[ \bar{\mu} \, \Psi_- \right]
    \>.
    \label{Wonshell2}
\end{equation}
Moreover, the following compatibility conditions hold:
\beq \label{compat}
\langle \Psi_\pm , dy^m \wedge \Psi_\mp \rangle = \langle \Psi_\pm , \iota_m  \Psi_\mp \rangle=0 \, .
\eeq
This allows us to take the warp factor out of the derivative in the first term on the RHS of \eqref{Wonshell2}, which then combines with the second term in \eqref{Wonshell1}. 
We are then left with 
\begin{eqnarray}
    \langle W_{\rm GCG} \rangle &=& 2 \pi i \int_{M_6} \<e^{3A-\phi} \Psi_- ,  
    d_H \left(e^{-\phi} \re \Psi_+\right) \>  - \frac{3 \pi i}{2}  \mu \int_{M_6} e^{2A-2\phi} \< \Psi_-,  \bar{\Psi}_- \> \nn \\
    &=& \left(16 - 12\right) \pi \mu  \int_{M_6}  e^{2A-2\phi} \mathrm{vol}_6 = \mu \, {\cal{N}}, 
    \label{byparts}
\end{eqnarray}
where in the first line we have integrated the first term by parts and used \eqref{susy1} together with  \eqref{purespinorVol}.  

For later reference, we note that, using properties of the Mukai pairing,  we can actually compute the first term in \eqref{byparts} exactly as above but without the need  to integrate by parts, i.e.~directly at the level of the integrand. This is because\footnote{Due to the compatibility condition \eqref{compat} and the fact that $\langle e^{B} \Psi_-,e^{B}\Psi_+ \rangle = \langle \Psi_-,\Psi_+ \rangle$ it is enough to consider the exterior derivative without the twisting by $H$, and without dilaton and warp factors. Then, we see that 
\begin{equation}
   \langle \Psi_-,d\Psi_+\rangle - \langle d\Psi_-,\Psi_+\rangle
   = - d[\Psi_-|_1\wedge \Psi_+|_4- 
   \Psi_-|_3\wedge \Psi_+|_2 + \Psi_-|_5\wedge \Psi_+|_0]=0    \, . 
\end{equation}
Here we have used that from the compatibility condition \eqref{compat} the five-form being differentiated on the LHS vanishes when it is wedged with any one-form and also when it is contracted with any vector, so it must be zero. This implies \eqref{parts}.} 
\begin{equation}
\label{parts}
    \<e^{3A-\phi} \Psi_-,   i d_H [ e^{-\phi} \mathrm{Re} \Psi_+]\> = \<d_H [e^{3A-\phi} \Psi_-],   i  e^{-\phi} \mathrm{Re} \Psi_+\>
    \, .
\end{equation}
We conclude that the integrand in \eqref{W10D} can be evaluated on-shell, giving 
\begin{equation}
\big\langle e^{3A-\phi} \Psi_-,F + i \, d_H \left( e^{-\phi} \re \Psi_+\right)\big\rangle = 4\mu e^{2A-2\phi} {\rm vol}_6 \,.
\label{WindetrandMU}    
\end{equation}
This will be useful in section \ref{sec:localised} below.

\subsection{SU(3) structure and Minkowski solutions}
\label{sec:GVW}
 
Configurations where $\eta^1$ and $\eta^2$ are parallel everywhere on the internal manifold up to a constant phase correspond to SU(3) $\subset$ O(6) structure compactifications. In these solutions the pure spinors reduce to\footnote{There is actually an overall phase in both pure spinors, given by the relative phase between the internal spinors: $\eta^1_+= ie^{ i\theta} \eta^2_\mp$. The relevant  sypersymmetry we use throughout this paper is the one compatible with O3- and O7-planes, namely $\theta=0$ \cite{Grana:2005jc}.} 
\begin{equation} \label{SU3ps}
    \Psi_-=  \Omega \ , \quad  \Psi_+= \exp (i J) \,, 
\end{equation}
where $J$ and $\Omega$ are a real (1,1)-form and a holomorphic (3,0)-form, respectively. The conditions in Eqs.~\eqref{compat} and \eqref{purespinorVol} then read 
\begin{equation}
J \wedge \Omega = 0 \,, \qquad \frac{1}{6} J\wedge J \wedge J = -\frac{i}{8}\Omega \wedge \bar{\Omega} = \mathrm{vol}_6\,. 
\end{equation}
Consequently, the superpotential \eqref{W10D} reduces to the usual Gukov-Vafa-Witten (GVW) expression \cite{Gukov:1999ya} 
\begin{equation}
W_{\mathrm{GVW}} = \pi \int_{M_6} e^{3A-\phi}  \Omega \wedge  G_3  \, . 
\label{GVW}
\end{equation}

It is not hard to see that, in this context, the supersymmetry equations reduce to the well-known type IIB supersymmetric Mink$_4$ solutions compactified on warped Calabi-Yau manifolds \cite{Grana:2000jj}.
Indeed, due to the absence of a 1-form component in $\Psi_-$, the 2-form component of Eq.\eqref{susy1} implies $\mu=0$, while the corresponding 4-form equation and the 3-form component of \eqref{susy2} read
\begin{equation} \label{SU3susy2}
    d\left(e^{3A-\phi} \Omega \right) = d\left(e^{2A-\phi} J\right) = 0. 
\end{equation}
Moreover, from the three-form components of \eqref{susy1} and \eqref{susy3} one also finds
\begin{equation}
      H \wedge \Omega =0\,, \qquad  e^{-\phi} H -  \tilde{F_3} = 0 \,, 
\end{equation}
so that $G_3$ must be ISD and its (0,3) component must vanish. 
The remaining equations give
\begin{equation}
    d\left(4A-\phi\right) = e^{\phi} \star_6{F}_5\,, \qquad
    d\phi  \wedge 
    J\wedge J = 
    -2 e^{\phi} \star_6  F_1 \ .
    \label{F1andF5Mink}
\end{equation}
By defining the relevant 5-form flux and the axio-dilaton as 
\begin{equation}
\label{F5AxioDilatonDef}
    F_5 = (1+*_{10}) \, \mathrm{vol}_4 
    \wedge d\alpha
    \, , \qquad  
    \tau = C_0 + i e^{-\phi} \,, 
\end{equation}
with $\alpha = \alpha(y)$, the conditions in \eqref{F1andF5Mink} can be written as 
\begin{equation}
    d\left(4A-\phi-\alpha
    \right) = 0 \,, \quad d\tau \wedge \Omega = 0\,.
    \label{F1andF5Mink2}
\end{equation}
Hence, one can have a non-trivial warp factor, related to the 5-form flux, while $\tau$ must be holomorphic.

\section{Revisiting the effect of the gaugino condensate} 
\label{sec:susywgauginos}

In this section we focus on the situation where one includes a stack of D-branes undergoing gaugino condensation, and discuss how such non-perturbative effects can be encoded in a set of modified supersymmetry conditions. In doing so, we combine the different elements considered originally in \cite{Koerber:2007jb,Dymarsky:2010mf} and more recently in \cite{Bena:2019mte,Kachru:2019dvo}, bringing them together into a consistent picture. 

\subsection{Supersymmetry conditions with localized terms}

We focus on the D7-brane case for concreteness, and because it is what we will be interested in in the following sections. 
The set of modified supersymmetry conditions we propose reads as follows: 
\begin{subequations}
\label{susymod}
 \begin{align}
d_H \left(e^{3A-\phi} \Psi_-\right) &= 2 i \mu e^{2A-\phi} \im \Psi_+  - 2 i \svev \delta^{(2)}\left[ \Sigma_4\right]\, , \label{susy1mod}\\
d_H \left(e^{2A-\phi} \im \Psi_+ \right) &= 0 \, ,  \label{susy2mod}\\
d_H \left(e^{4A-\phi} \re \Psi_+ \right) &= 3 e^{3A-\phi} \re \left[ \bar{\mu} \, \Psi_- \right] + e^{4A} *_6 \alpha( F) -  e^{A} \delta^{(0)} \left[ \Sigma_4\right]\re \left[\bar{\svev} \Psi_-\right] \, . \label{susy3mod}
 \end{align}
\end{subequations} 
Here  $\delta^{(2)}\left[ \Sigma_4\right]$ is the localized 2-form Poincar\'e dual to the four-cycle wrapped by the branes, namely for any closed 4-form $\omega_4$ one has 
\beq \label{delta2}
\int_{M_6} \omega_4 \wedge \delta^{(2)}\left[ \Sigma_4\right] = \int_{\Sigma_4} \omega_4  \,,
\eeq
while $\svev$ is the VEV of the usual condensate superfield, related to the gaugino bilinear by 
\begin{equation}
    \svev = \frac{1}{16 \pi^2} \langle \lambda \lambda \rangle \, .
    \label{svev}
\end{equation}
Moreover, $\delta^{(0)}[\Sigma_4]$ is the scalar version of the delta function, defined as \cite{Grana:2020hyu}
\beq \label{delta0calibrated}
\delta^{(0)}[\Sigma_4]= (\im \Psi_+)^{(2)} \cdot \delta^{(2)}[\Sigma_4] \quad \Rightarrow \quad \delta^{(0)}[\Sigma_4] \, {\rm vol}_6 = \< \re \Psi_+, \delta^{(2)}[\Sigma_4]  \> \, .
\eeq
The analysis for other types of branes and for the type IIA case is analogous. For gaugino condensates on
other type IIB Dp-branes wrapping $p-3$ cycles, one replaces  $\delta^{(2)}(\Sigma_4)$ by $\delta^{(9-p)} (\Sigma_{p-3})$. For the type IIA cases one further exchanges $\Psi_+$ with $\Psi_-$.

\subsection{Motivation}
\label{sec:Wvevgaugino}

We now explain how this proposal comes about. 
Let us start with the modification to the first supersymmetry equation, namely Eq.~\eqref{susy1mod}. 
In the ten-dimensional language, this should come from the F-term condition associated with the variation of the superpotential with respect to the superfield $T$, defined in \eqref{Z,T}. As advocated in \cite{Koerber:2007xk,Dymarsky:2010mf} and further discussed in \cite{Bena:2019mte}, assuming a non-trivial gaugino condensate on a stack of calibrated D7-branes leads to an extra contribution to the F-term. Indeed, in the 4D ${\cal{N}}=1$ superspace description of the Yang-Mills (YM) theory living on the branes one must include a chiral contribution to the effective Lagrangian of the form  
\begin{equation}
  \frac{i}{8\pi} \int d^2\theta \, \tau \Tr \left[W^\alpha W_\alpha\right], 
\end{equation}
where $\tau$ is the complexified gauge coupling\footnote{We use the same notation as for the axio-dilaton defined in Eq.~\eqref{F5AxioDilatonDef}. The distinction should be clear from the context.} and $W_\alpha$ is the usual chiral superfield, i.e.
\begin{equation}
    \tau = i \frac{4\pi}{g_{\rm YM}^2} +   \frac{\theta_{\rm YM}}{2\pi}\ , \quad \ W_\alpha = 
    -i \lambda_\alpha + \cdots\, .
\end{equation}
The non-perturbative effects that generate a non-trivial expectation value for the condensate superfield 
\begin{equation}
    S = \frac{1}{16\pi^2} \Tr \lambda^\alpha \lambda_\alpha 
\end{equation}
are then captured by the Veneziano-Yankielowicz (VY) superpotential\footnote{Here we take the gauge group to be SU(N) for simplicity. When considering, say, D7-branes coincident with O7-planes such that the charges are cancelled locally, it should be taken to be SO(8) instead. This introduces only minor modifications. } \cite{Veneziano:1982ah}
\begin{equation}
     \quad \ W_{\rm VY} = W_0 + 2\pi i \tau S + NS\left[1-\log \left(S/\mu_0^3\right)\right]\, ,
    \label{WVY}
\end{equation}
where $\mu_0$ is the scale at which $\tau$ is defined, and $W_0$ is taken to be independent of $S$. 

As the effective four-dimensional YM coupling comes from integrating over $\Sigma_4$, it depends on its volume, and also on the RR potentials involved. More precisely, we have 
\begin{equation}
    \tau = \int_{\Sigma_4} \left(C + i e^{-\phi}
    \re \Psi_+\right)|_{\Sigma_4} = \int_{M_6} \langle T,- \delta^{(2)}(\Sigma_4)\rangle \,, 
    \label{tau4cycle}
\end{equation}
where we have used that $T$ is the calibration form on the holomorphic cycle $\Sigma_4$, which defines the associated volume form and Chern-Simons coupling. Therefore, $\tau$ must be seen as a function of the chiral field $T$. The corresponding F-term condition then picks up an extra contribution given by the last term on the RHS of \eqref{susy1mod}. The exterior derivative of the resulting condition still gives \eqref{susy2mod}, so it is not modified. Note that upon integrating out $S$ in Eq.~\eqref{WVY} one finds the VEV and effective superpotential 
\begin{equation}
    \langle S\rangle  = \mu_0^3 \exp \left(\frac{2\pi i \tau}{N} \right) \, , \quad 
    W_{\rm eff} = W_0 + N \langle S\rangle \, ,
\end{equation}
used in the 4D EFT analysis of \cite{Kachru:2003aw}. 

\medskip

On the other hand, the argument for the third supersymmetry condition in Eq.~\eqref{susy3mod} is more delicate. Indeed, even in the absence of a gaugino condensate it is not straightforward to see that this equation is equivalent to the F-term condition for the chiral field $Z$, as it was discussed in \cite{Koerber:2007xk}. Alternative arguments supporting similar modifications were presented in \cite{Dymarsky:2010mf,Kachru:2019dvo} assuming an expansion in powers of the gaugino condensate VEV, while the authors of \cite{Dymarsky:2010mf} further work in the rigid, decompactified limit. We now argue that no approximations are needed to motivate  Eq.~\eqref{susy3mod} in generalized geometry compactifications involving non-perturbative sources. 

For that, consider the DBI action describing the D7-brane theory. As it was shown recently in \cite{Grana:2020hyu} (see also \cite{Lust:2008zd}), for a generic internal manifold in a GCG compactification, the quadratic terms in the gaugino action can be written as 
\beq \label{4daction}
S_{\lambda\lambda}=\int d^4x \, \left( \frac{i}{2}f \, \bar{\lambda}_+\gamma^\mu\nabla_\mu\lambda_+ + \frac{1}{2} m_{\lambda} \,  
\bar{\lambda}_-\lambda_+ + c.c.   \right)  \, ,
\eeq
with $\bar{\lambda}_- \lambda_ +  = i 16 \pi^2 \bar{S} $, and where\footnote{Here we have included a warp factor missing in the original version of \cite{Grana:2020hyu}.} 
\begin{equation}
    m_\lambda = -\frac{i}{8\pi} \int_{M_6} \delta^{(0)} [\Sigma_4] \,  e^{A} 
    \big\langle \Psi_-,F + i \, d_H \left( e^{-\phi} \re \Psi_+\right)\big\rangle\,.
    \label{gauginomass1}
\end{equation}
Note that the integrand is precisely that of the superpotential,  Eq.~\eqref{W10D}\footnote{Although this holds for the D7 case, for other D-branes one must be more careful when considering the contribution of the NSNS 3-form flux \cite{Grana:2020hyu}. This is in agreement with the results of \cite{Dymarsky:2010mf}.}. Focusing on the RR flux contribution to the gaugino mass, we have 

\begin{equation}
  S_{\lambda\lambda,F}=  2 \pi   \int_{M_6} e^{A} \delta^{(0)} \left[ \Sigma_4\right]
   \langle \re \left[\bar{\svev} \Psi_-\right],F
   \, \rangle \,.
\end{equation}
This additional term in the bulk action provides a new source in the equations of motion for the RR fluxes. More explicitely, Eq.~\eqref{EOMF} is modified as follows: 
\begin{equation}
\label{EOMFmod}
    d_H \left[e^{4A} *_6 \alpha (F)\right] = 
    d_H \left[e^{A} \delta^{(0)} \left[ \Sigma_4\right]
   \re \left(\bar{\svev} \Psi_-\right)\right]
    \,.  
\end{equation}
As discussed in Sec.~\ref{sec:susywogauginos} above,  we expect this to follow from the derivative of the third supersymmetry condition. We find that the localized term introduced in Eq.~\eqref{susy3mod} ensures that this is indeed the case. 

Importantly, Eq.~\eqref{susy3mod} should also account for an electric source for $H$ as implied by the corresponding contribution to the mass term in \eqref{gauginomass1}. Since $*_\alpha \Psi_- = i \Psi_-$, by acting with $*_\alpha$ on \eqref{susy3mod}  we see that the new term becomes  proportional to $\im \left[  \bar{\svev} \Psi_- \right]$. Further multiplying by $e^{-\phi}$ and taking the exterior derivative we find that the proposed non-perturbative correction is consistent with the coupling between $H$ and the gaugino condensate in Eq.~\eqref{gauginomass1}.   

\subsection{Self-consistency and interpretation}
\label{sec: consistency loc}

We now show that the system of equations given in \eqref{susymod} is self-consistent, and argue that the GCG superpotential encodes all ingredients relevant to the effective action, including non-perturbative terms. 

Given a solution to the modified supersymmetry conditions \eqref{susymod}, we can compute the on-shell value of the GCG superpotential defined in \eqref{W10D}. The procedure is analogous to what we described in Sec.~\ref{sec:Wvev}, except that we now get two additional contributions coming from the localized terms present in the first and third supersymmetry conditions. One comes from inserting the on-shell value of $\tilde{F}$ as given by \eqref{susy3mod}, similarly to \eqref{Wonshell2}, while the other comes from the integration by parts and the use of \eqref{susy1mod}, as was done in \eqref{byparts}. 
These additional contributions are given by 
\begin{equation}
\langle W_{\rm GCG}^{\rm loc}\rangle =-4 \int_{M_6} \< \svev \delta^{(2)}\left[ \Sigma_4\right],  
   e^{-\phi} \re \Psi_+\> + i \int_{M_6} \<e^{3A-\phi} \Psi_-, 
    e^{-3A} \delta^{(0)} \left[ \Sigma_4\right]\re \left[\bar{\svev} \Psi_-\right]
    \> \, ,
    \label{Wonshelllocterms}
\end{equation}
where we have used \eqref{delta0calibrated}.
Taking into account the definition of the scalar delta function in \eqref{delta0calibrated}, we find that the two terms in Eq.~\eqref{Wonshelllocterms} precisely cancel each other, namely 
\begin{equation}
\langle W_{\rm GCG}^{\rm loc}\rangle=0 \ .
\end{equation}
Although this cancellation was recently obtained in an extended version of appendix A in \cite{Kachru:2019dvo}, we believe that its origin has not been fully clarified. We stress that from the point of view adopted in the previous sections it is surprising to learn that the constant $\mu$ appearing in the supersymmetry equations corresponds to the on-shell value of the GCG superpotential \eqref{W10D} \textit{even in the presence of the gaugino condensate.} In other words, the \textit{explicit} contributions associated with the non-zero VEV of the gaugino bilinear vanish. 
There are, however, \textit{implicit} contributions since the on-shell values of the different fields -- and in particular that of the cosmological constant -- are indeed affected by the presence of the condensate. This is in contrast to the naive expectation according to which 
one should have 
$W \sim W_0 + W_{\rm np}$, where $W_0$, i.e.~the "flux superpotential",  would correspond to $W_{\rm GCG}$ evaluated on-shell, while $W_{\rm np}$ would in turn be associated to the VEV of a putative additional non-perturbative term in the "full superpotential".

This further agrees with what one finds both in the heterotic context  \cite{Frey:2005zz,Minasian:2017eur,Held:2010az} and (very similarly) in type I theories, although of course in these cases the gaugino condensate is not localised. 
Nevertheless, it would be reassuring to understand exactly how $W_{\rm GCG}$ as defined in \eqref{W10D} is able to fully capture the backreaction associated to gaugino condensation on the localized D7-branes. In other words, we would like to understand how the non-perturbative terms in $W_{\rm VY}$ are generated, see Eq.~\eqref{WVY}. Although we have not been able to show this fully explicitly, we  suggest a possible mechanism for how this might happen.

Let us first recall how open string degrees of freedom are captured by $W_{\rm GCG}$ in GCG compactifications, as it was discussed in section 3.3 of \cite{Koerber:2007xk}. 
Including D-branes (or O-planes) in a given setting induces a localized source term $\delta_{\rm Dp}$ in the Bianchi identities for the RR fluxes, see Eq.~\eqref{Bianchi}. Locally, we can write $\delta_{\rm Dp} = d_H \theta_{\rm Dp}$ for some $\theta_{\rm Dp}$. Then, we can formally split the physical RR fluxes according to $F = F_{0} + \theta_{\rm Dp}$. This distinguishes two contributions to the superpotential, namely 
\begin{equation}
    W_{\rm GCG} =  \int_{M_6} \<e^{-B}Z, G\> =  \int_{M_6} \<e^{-B}Z, G_{0}\> +  \int_{M_6} \<e^{-B}Z, \theta_{\rm Dp}\> = W_{0} + W_{\rm op}\,,
    \label{Wclosedopen}
\end{equation}
where $G$ was defined in \eqref{defG}, while $G_0$ is defined analogously with the replacement $F\to F_0$. One finds that $W_{\rm op}$ computes the open-string superpotential of 
\cite{Martucci:2006ij}. Conversely,  $W_{0}$ is interpreted as the closed-string superpotential. We see that both of them come from $W_{\rm GCG}$.

We propose that a similar phenomenon occurs when the gaugino bilinear on a stack of D-branes acquires a non-zero expectation value. Similarly to the D-branes themselves sourcing bulk RR fluxes, it was argued in \cite{Dymarsky:2010mf} that gaugino condensates constitute sources for the geometry itself. More precisely, they source the degrees of freedom contained in the holomorphic variable $Z$. This is because, as reviewed above, the periods of  $T$ define the effective coupling of the gauge theory on stacks of space-filling D-branes wrapping calibrated internal cycles \cite{Koerber:2010bx}. For instance, a generic Mink$_4$ vacuum involving such non-perturbative contributions should satisfy a supersymmetry condition of the form $ d_H \left(e^{-B}Z\right) = \delta_{\rm np}$, where $\delta_{\rm np}$ is the localized non-perturbative current proportional to the gaugino condensate. Hence, at least locally, we can define some $\theta_{\rm np}$ for which $ d_H \theta_{\rm np} = \delta_{\rm np}$. We interpret this as capturing the backreaction of the geometry, and split $e^{-B}Z = e^{-B}Z_{0} + \theta_{\rm np}$. Combining this with the discussion above, the superpotential reads 
\begin{equation}
    W_{\rm GCG} =  \int_{M_6} \<e^{-B}Z, G\> =  W_{0} + W_{\rm np} \,, 
\end{equation}
where now $W_{\rm np}$ contains all terms proportional to the condensate, while the remaining ones are packed into $W_0$. The former containts two contributions, $W_{\rm np} = W_{\rm np,1} + W_{\rm np,2}$, the first of which is given by
\begin{equation}
    W_{\rm np,1} = \int_{M_6} \<\theta_{\rm np}, G_0 \>  =\int_{M_6} \<d_H\theta_{\rm np}, C_0+i e^{-\phi} \re \Psi_+\> = 2\pi i \tau S \,. 
    \label{Wnp1}
\end{equation}
where in the second equality we have integrated by parts. Hence, $W_{\rm np,1}$ gives precisely the second term in the VY superpotential \eqref{WVY}. 
We conclude that at least part of the non-perturbative terms in the superpotential are indeed generated upon evaluating $W_{\rm GCG}$ on-shell, and  propose that the rest of $W_{\rm VY}$ is generated as well, arising as from the  combination of the two effects we have just described. Indeed, the second contribution comes from implementing the replacement $e^{-B}Z \to  e^{-B}Z_{0} + \theta_{\rm np}$  \textit{inside} the open string superpotential $W_{\rm op}$, i.e.~the final term in \eqref{Wclosedopen}. This leads to 
\begin{equation}
    W_{\rm np,2} =  \int_{M_6} \<\theta_{\rm np}, \theta_{\rm Dp}\> \,. 
    \label{Wnp2}
\end{equation}
Although we have not been able to evaluate this explicitly, we note that it must be proportional to both the number of branes $N$ and the condensate $S$. This matches our expectation for the final term in $W_{\rm VY}$ as defined in Eq.~\eqref{WVY}. However, it would be nice to understand exactly how the $\log S$ factor appears. 

Our proposal is further motivated by the well-studied case of geometric transitions in conifolds, which can be described in the GCG language applied to the SU(3) structure case.
Consider, for concreteness, $N$ D5-branes wrapping the compact two-cycle $\Sigma_2 = S^2$ located down the throat of the resolved conifold.  At large $N$, one has a geometric transition where the 2-cycle shrinks to zero size, and a 3-cycle opens up, thus leading to the deformed conifold geometry \cite{Cachazo:2001jy,Atiyah:2000zz,Maldacena:2009mw}. The original D5-branes disappear, and we are left with fluxes together with a modified geometry.
After the transition, we can evaluate the superpotential as 
\begin{equation}
    W = \int_{M_6} \Omega \wedge (F_3 + i e^{-\phi} d J) =  \int_{S^3} \Omega \int_{B_3} (F_3 + i e^{-\phi} d J) - 
    \int_{B_3} \Omega \int_{S^3} (F_3 + i e^{-\phi} d J)\,, 
    \label{WD5conifold}
\end{equation}
where $B_3$ is the non-compact 3-cycle dual to the $A$-cycle $S^3$. 
Now, the integral of $\Omega$ on the resulting $S_3$ is set by $\svev$, while $B_3$ can be thought to have a boundary given by an $S^2_c$ at a radial cutoff scale $\Lambda_c$. Hence, the first term on the RHS of \eqref{WD5conifold} gives the second term in $W_{\rm VY}$, namely
\begin{equation}
     \int_{S^3} \Omega \int_{B_3} (F_3  + i e^{-\phi} d J) \sim   S \int_{S^2_c} (C_2 + i e^{-\phi} J) =  2\pi i \tau S\,, 
    \label{defConifoldT1}
\end{equation}
up to an overall constant, where $\tau$ corresponds to the running YM coupling. 
On the other hand, evaluating the second term with the help of the explicit solution presented in  \cite{Maldacena:2000yy}, and analogously to what was done in \cite{Cachazo:2001jy} (see also \cite{Maldacena:2009mw}) one gets
\begin{equation}
     - \int_{B_3} \Omega \int_{S^3} (F_3 + i e^{-\phi} d J) \sim
      NS\left[1-\log \left(S/\Lambda_c^3\right)\right], 
      \label{defConifoldT2}
\end{equation}
which, together with \eqref{defConifoldT1} reproduces the full VY superpotential, as expected. 
Conversely, it was proposed in \cite{Dymarsky:2010mf} that one can consider the same computation \textit{before} the geometric transition (or more generally at smaller values of $N$ such that the transition is not induced), so that the $S^3$ cycle is trivial in homology but the $S^2$ is not. Including the effect of gaugino condensation in an SU(3) structure background  one has\footnote{The first equation is simply the flat limit of \eqref{susy1mod} with  constant warp factor and dilaton.} 
\begin{equation}
   d \Omega = 2i \svev   \delta^{(4)}[\Sigma_2]\,, \quad 
     d F_3 = -N \delta^{(4)}[\Sigma_2] \ , \quad 
    \label{dFdOMD5}
\end{equation} where we have set $H=0$.
The evaluation of the GCG superpotential in this context then ammounts to a computation very similar to what we have described above in Eqs.~\eqref{Wnp1} and \eqref{Wnp2}. Since we expect to find the same result, namely $W_{\rm VY}$, we consider this as an example of the general mechanism proposed there. 



\section{Smearing the condensate: KKLT as a proof of concept}
 \label{sec:smeared}
 
The modified supersymmetry conditions \eqref{susymod} imply that for localised sources a 10D description of the KKLT solution can not have SU(3) structure. For instance, this can be deduced from the 2-form component of \eqref{susy1mod} \cite{Koerber:2008sx,Dymarsky:2010mf,Bena:2019mte}. One thus needs to consider internal manifolds with what is known as a dynamical SU(2) structure group. However, explicit models of this type are hard to construct in practice (see for instance \cite{Heidenreich:2010ad}). 

There are (at least) two ways of evading these difficulties. On the one hand, one could study these solutions as small perturbations (in $\svev$, or equivalently, in the deviation from SU(3) structure) on top of the flat solution. This was attempted in \cite{Dymarsky:2010mf,Kachru:2019dvo}. On the other hand, as is often done in the context of string compactifications, one can try to simplify the problem by \textit{smearing} the source. This possibility was suggested in \cite{Koerber:2008sx}, although at the time the modified version of the third supersymmetry equation \eqref{susy3mod} was not available. In this section we reconsider this proposal. By carefully carrying out the smearing procedure for the gaugino condensate, we find that the modified supersymmetry equations given in \eqref{susymod} lead to a remarkably simple solution. The latter is such that SU(3) structure is maintained, and turns out to be in perfect agreement with the original effective four-dimensional analysis by KKLT \cite{Kachru:2003aw}.



Let us see how this works. So far we have implicitly assumed that the gauge fluxes ${\cal{F}}$ on the worldvolume of the D7-branes vanish, which allowed us to think of the source in \eqref{susy1mod} as a two-form. (Otherwise we would have needed to include higher degree forms coming from $ \delta^{(2)}\wedge e^{\cal{F}}$). This was done also because ${\cal{F}}$ was explicitly set to zero when computing the different contributions to the gaugino mass terms in \cite{Grana:2020hyu}, which motivated our modification of the other supersymmetry condition. The appropriate smearing is given by the replacement   
\begin{equation}
\label{smearingdelta2}
    \delta^{(2)}[\Sigma_4] \to \gamma e^{2A-\phi} J \,,
    \qquad \gamma = - \frac{4\pi\sigma_4}{3{\cal{N}}} \,,
\end{equation}
where ${\cal{N}}$ was defined in \eqref{Ndef}, while $\sigma_4$ keeps track of the volume of $\Sigma_4$. The numerical constant $\gamma$ is fixed by requiring that the integral of the localized source and that of its smeared counterpart give the same result, namely
\begin{equation}
    \sigma_4 = \int_{M_6} \<
    e^{-\phi} \re \Psi_+ , -\delta^{(2)}[\Sigma_4]
    \> = - \int_{M_6} \<
    e^{-\phi} \re \Psi_+ , \gamma e^{2A-\phi} J
    \> 
    = -\frac{3 \gamma {\cal{N}}}{4\pi} \,,
\end{equation}
where we have used \eqref{purespinorVol}.
The same can be done for the scalar delta function. We set
\begin{equation}
    \delta^{(0)}[\Sigma_4] \to 3\gamma e^{2A-\phi}
    \label{smearingdelta0}
\end{equation}
so that, using \eqref{delta0calibrated}, we get 
\begin{equation}
    \sigma_4 = -\int_{M_6} \delta^{(0)}[\Sigma_4] \, e^{-\phi}{\rm vol}_6 
    = - \frac{3 \gamma {\cal{N}}}{4 \pi}\, ,
\end{equation}
as expected. The fact that the coefficient appearing in the smearing of the scalar delta function is three times that of the localized 2-form delta is consistent with the identity \cite{Baumann:2010sx}
 \begin{equation}
     \frac{1}{2}J\wedge J\wedge \delta^{(2)} = \frac{1}{6} J\wedge J\wedge J \delta^{(0)} \,.
 \end{equation}
 
Inserting \eqref{smearingdelta2} into Eq.~\eqref{susy1mod}, we find that, as anticipated above, we do not need to consider an internal manifold with a more general structure group than SU(3). Indeed, the problematic two-form component now reads
\begin{equation}
    d(e^{3A-\phi}\Psi_-|_1) = 2i (\mu - \gamma \langle S \rangle ) e^{2A-\phi}J \,,  
\end{equation}
which is satisfied when $\Psi_-$ has no 1-form component provided 
\begin{equation}
    \mu = \gamma \svev \,.
    \label{muS}
\end{equation}
Moreover,  we get 
\begin{equation} \label{SU3susy2b}
    d\left(e^{3A-\phi} \Omega \right) = d\left(e^{2A-\phi} J\right) = 0 \,. 
\end{equation}
Although this is starting to sound very similar to the configuration described in Sec.~\ref{sec:GVW}, there are some crucial differences. First, the extended part of the solution is now AdS$_4$. The presence of $\mu$ generates a non-trivial contribution in the RHS of the 6-form component of  Eq.~\eqref{susy1mod}. This reads  
\begin{equation}
    H \wedge \Omega =   \frac{ \mu}{3} e^{-A} J\wedge J \wedge J = -
     \frac{\mu}{4} \, 
    e^{-A} \bar{\Omega}
    \wedge \Omega \,.  
    \label{HOMsmeared}
\end{equation}
Hence, we find that an additional (0,3) component in the 3-form flux is needed. 

Notably the condition \eqref{HOMsmeared} is the only place where the terms proportional to the cosmological constant in the system of equations \eqref{susy1mod} do not cancel with those coming from the smeared gaugino condensate sources. Indeed, the replacement of \eqref{smearingdelta0}, including the crucial factor of $3$, implies that the first and third terms on the RHS of the polyform equation \eqref{susy3mod} cancel exactly upon imposing \eqref{muS}. Consequently, the ISD condition on the $G_3$ fluxes is preserved! From \eqref{HOMsmeared} we get
\beq  \label{03pieceH}
H_{(0,3)}=-\frac{1}{2} \, e^{-A}\,  {\rm Re}(\bar \mu \Omega) \,,  \qquad F_{(0,3)} = -\frac12 e^{-\phi -A} {\rm Im}(\bar \mu \Omega) \,. 
\eeq
Note that the phase of the gaugino condensate sets the phase of the cosmological constant, which then fixes the phase of these flux terms relative to $\Omega$. This suggests that the cycles dual to the NSNS and RR three-form fluxes are Special Lagrangian. 

By using \eqref{SU3susy2b}, we find that the Bianchi identities for these fluxes are satisfied iff 
\beq
d(4A-\phi)=0 \ , \qquad d \tau \wedge \Omega=0 \ . 
\label{d4Aphidtau}
\eeq
As the supersymmetry equations involving the relevant RR fluxes are the same as in section \eqref{GVW}, holomorphicity of the axio-dilaton is consistent with the 5-form component of Eq.~\eqref{susy3mod} in the smeared approximation. On the other hand, the corresponding 1-form implies that in order to satisfy \eqref{d4Aphidtau} we must have $F_5=0$. %
The Bianchi identity for $F_5$ then reads
\beq
H_3 \wedge F_3 = \delta_{\rm D3}\,,
\eeq
where $\delta_{\rm D3}$ stands for any source with D3-charge, implying that these must be smeared as well. This Bianchi identity then turns into the tadpole cancellation condition.  

\subsection{Summary and effective theory}

In summary, we see that not much has changed as compared to the supersymmetric Mink$_4$ solutions described in section \ref{sec:GVW}. Upon including the gaugino condensate sources as in \eqref{susymod} and smearing them according to Eqs.~\eqref{smearingdelta2} and \eqref{smearingdelta0}, we have constructed supersymmetric AdS$_4$ solutions which have the following characteristics: 
\begin{itemize}
    \item the internal manifold is still K\"ahler with an SU(3) structure group, and for constant dilaton and warp factor it is still CY,  
    
    \item the axio-dilaton is still holomorphic, 
        
    \item the 3-form flux $G_3$ is still ISD, and its primitive (2,1) component sets the mass scale for complex structure and axio-dilaton deformations, 
    
    \item the 5-form flux $F_5$ must now vanish, which also sets $4A=\phi+{\rm const}$, i.e.~the Einstein frame warp factor $A_{\rm E}$ is constant, 

    \item the value of the cosmological constant, encoded in $\mu$, is fixed by that of the gaugino condensate $\svev$ as in Eq.~\eqref{muS}, where the coefficient $\gamma$ is fixed by the consistency of the smearing approximation, and 
    
    \item the 3-form flux $G_3$ now acquires a (0,3) piece proportional to the cosmological constant, see \eqref{03pieceH}.
\end{itemize}

We now compare with the effective theory discussed in \cite{Kachru:2003aw}.
In terms of the corresponding on-shell superpotential, the relation between the curvature scale $\mu$ appearing in the supersymmetry conditions and the gaugino condensate derived in \eqref{muS} becomes   
\begin{equation} \label{KKLT10D}
    \langle W_{\rm 4D}\rangle = \mu \, {\cal{N}} =  - \frac{4\pi\sigma_4}{3}\svev =  - \frac{4\pi\sigma_4}{3N} \langle W_{\rm np} \rangle \,, 
\end{equation}
where we have used Eq.~\eqref{smearingdelta2} together with $\langle W_{\rm np} \rangle = N \langle S \rangle$. This precisely reproduces the KKLT results \cite{Kachru:2003aw}. 

The exact matching between our ten-dimensional smeared solution and the effective theory we have obtained is, in some sense, not entirely surprising. Indeed, it is consistent with the expectation that the latter captures the physics of the zero-modes on the internal manifold. This is exactly the sector of the theory we have restricted to when carrying out the smearing procedure. This is similar to what happens in the DGKT case in type IIA \cite{DeWolfe:2005uu,Marchesano:2020qvg}. As in that case, we also find that a specific combination of the warp factor and the dilaton must be constant in the smeared limit.  

Moreover, we also confirm the interpretation of \cite{Kachru:2003aw}: the non-vanishing cosmological constant originates from the presence of ISD supersymmetry-breaking fluxes and non-perturbative physics captured by gaugino condensation. In our construction, their precise balance is showcased in Eq.~\eqref{HOMsmeared}. 

On the other hand, note that by looking at the on-shell value of the (0,3) fluxes given in \eqref{03pieceH} we  \textit{can not} isolate a term independent of the K\"ahler modulus $\sigma_4$ contributing to the superpotential (this was denoted $W_0$ in \cite{Kachru:2003aw}). Indeed, the condensate itself is expected to source (0,3) 3-form flux. This was shown in \cite{Baumann:2010sx,Dymarsky:2010mf}, where this component was found to be completely localized on $\Sigma_4$. Upon smearing, this becomes an extra contribution to the total $G_{(0,3)}$. In this sense, one should not think about the supersymmetric AdS KKLT vacua as a two-step procedure, the first involving susy-breaking fluxes and the second introducing the gaugino condensate.
These two ingredients come hand in hand in order to produce a stable supersymmetric solution.




\subsection{Scale separation}

Here we discuss whether the smeared solution allows for scale separation. For that, we consider the scalings of the various fields that leave the (smeared) supersymmetry equations invariant. There are two variables in the game: $g_s = e^{\phi}$ and $R$, the characteristic scale of the compactification, assuming there is only one such scale. Under these assumptions, the $p$-form components of the pure spinors scale as $\sim R^{p}$, namely
\beq
\Omega \sim R^3 \,, \quad J \sim R^2 \, , \quad  \Rightarrow   \quad \sigma_4 \sim \frac{R^4}{g_s}
\, .
\eeq
The coefficient $\gamma$ defined in \eqref{smearingdelta2}, which relates the condensate to the cosmological constant by \eqref{muS}, scales as
\beq
\gamma\sim \frac{\sigma_4}{\cal N} \sim \frac{g_s}{R^2}\, ,  
\eeq
hence the cosmological constant scales as
\beq
\mu \sim \svev \frac{g_s}{R^2}\, .
\eeq
From the ISD condition we have $e^{-\phi}H = \tilde{F}_3 \sim F_3$. On the other hand, \eqref{03pieceH} implies that the (0,3) component of $H$ scales as
\beq
H_{(0,3)} \sim \mu R^3 \sim \svev {g_s} R \, \quad \Rightarrow \quad  F_{(0,3)} \sim \svev R \, ,
\eeq
where we have assumed that $e^A\sim 1$.
The (2,1) components are on the contrary not related to the gaugino condensate expectation value, so that  the full $G_3$ flux scales as
\beq
G_3=G_{(2,1)}+G_{(0,3)} \sim G_{(2,1)} + \svev R .
\eeq
The tadpole cancellation condition then works as follows
\begin{align} \label{tadpole}
  \frac{1}{24}\chi \left(X_4\right) &= 
    N_{\rm D3} +  \int_{M_6} H \wedge F_3 \nonumber\\
    &= N_{\rm D3} 
     -2 i g_s \int_{M_6} G_{(2,1)} \wedge \bar G_{(2,1)} 
     -2i g_s \int_{M_6}G_{(0,3)} \wedge \bar G_{(0,3)} \\ & \sim N_{\rm D3} + g_s \, G_{(2,1)}^2  +  g_s \svev^2 R^2 \, .
     \nonumber    
     \end{align}
 Here $\chi(X_4)$ is the Euler characteristic of the elliptic CY 4-fold of the associated F-theory compactification, which can take values from ${\cal O}(100)$ to ${\cal O}(10^6)$ \cite{Gray:2014fla,Candelas:1997eh}. One might think that by flux quantization, both contributions to the tadpole coming from the fluxes have to be of a similar order, but recall that flux quantization applies to real cycles, while these cycles are complex. In other words the split of the integer value of the flux induced charge into the two terms in \eqref{tadpole} depends on the complex structure moduli.   

The question is what sets the value of the gaugino condensate, which is related to the cosmological constant
via \eqref{muS}. Once this relation is plugged in, the terms involving the condensate cancel with the cosmological constant ones. Consequently, geometric quantities do not rescale with $\svev$. Such a relation comes only from the non-perturbative superpotential, namely the relevant terms in Eq.~\eqref{WVY}. On-shell, this leads to an exponential behaviour of the type
\begin{equation} \label{svevvolume}
    |S| \sim e^{-\frac{1}{g_s} R^4} \ .
\end{equation}
  
Plugging this back into the tadpole cancellation condition, we find that there is no  apparent contradiction. If one can attain the regime of large $R$ and small $g_s$, the contribution from the (0,3) piece is much smaller than the one from the (2,1), the latter giving the main contribution to the tadpole.
Everything then is consistent with the following scalings 
\begin{equation}
    \mu \sim \frac{g_s}{R^2}\, e^{-\frac{1}{g_s}R^4} \ , \quad  H_{(0,3)} \sim R \, e^{-\frac{1}{g_s}R^4} \, , \quad
   H_{(2,1)} \sim R^0 \ , \quad \Omega \sim R^3 \ , \quad J \sim R^2 \, ,  
\end{equation}
which imply
\begin{equation}
    \frac{\ell_{KK}}{\ell_{AdS}} = \mu{R} \sim \frac{g_s}{R} \, e^{-\frac{1}{g_s}R^4} \,.
\end{equation}
Hence, at least in the smeared solution, the AdS and KK scales are exponentially separated. This can be achieved if one can find quantized fluxes such that $W_0$ is very small. This was achieved recently in a family of explicit examples described in  \cite{Demirtas:2019sip}, where the authors provide flux configurations where the (0,3) pieces vanish in the limit where the prepotential has only polynomial terms, but they are non-trivial when the corresponding exponentially small corrections are included\footnote{Note, however, that such construction has been criticised in \cite{Lust:2022lfc}.}.

\section{Features of the localized solution}
\label{sec:localised}

We finish by providing some insights on the main features of any putative solution to the supersymmetry conditions \eqref{susymod} with localized sources, leaving the construction of the full configuration for future work.
We also settle the issue of divergencies and four-fermion terms analyzed in \cite{Hamada:2018qef,Kallosh:2019oxv,Hamada:2019ack,Hamada:2021ryq,Kachru:2019dvo}. 

\subsection{Dynamic SU(2) structure and IASD fluxes}


As stated in \cite{Koerber:2007xk,Bena:2019mte}, the modified conditions \eqref{susymod} imply that in order to find supersymmetric AdS$_4$ solutions sourced by a gaugino condensate on a stack of D7-branes one needs to leave the realm of SU(3) structure compactifications\footnote{This is easy to see from the two-form piece of \eqref{susy1mod}. For SU(3) structure, $\Psi_-$ is a three-form, therefore the left-hand side has no two-form piece, while both terms on the right hand side do. For the smeared solution these two terms cancel each other, but in the localised one this is no longer possible. Thus, a solution to this equation would require $\Psi_-$ to contain a one-form piece as well.}. Nevertheless, constructing explicit solutions with so-called "dynamic SU(2) structure" (where the alignment of the spinors $\eta^1$ and $\eta^2$ in the pure spinors \eqref{purespinors} depends on the position)    constitutes a considerable challenge.  

Some steps in this direction were given in \cite{Heidenreich:2010ad,Dymarsky:2010mf,Kachru:2019dvo}, see also \cite{Baumann:2010sx}. In \cite{Heidenreich:2010ad} the authors attempted to construct a solution of this type, and although the localized sources were not included explicitly, they managed to describe the region of the geometry close to the D7-branes wrapping the four-cycle at the bottom of a resolved $\mathbb{P}^2$ cone.
On the other hand, \cite{Dymarsky:2010mf} and \cite{Kachru:2019dvo} considered an expansion in powers of the (absolute value of the) gaugino condensate, and studied the solution at first order. In this regime, also the cosmological constant and the angle parameterising the deviation from SU(3) structure can be considered small. This approximation is expected to work best far away from the localized sources. It was argued in \cite{Bena:2019mte} that K\"ahler moduli stabilization can be understood from a consistent matching of both regimes of the solution. 

As expected, the gaugino condensate not only backreacts on the geometry, but it also affects the three-form fluxes. More precisely, and consistent with our discussion around Eq.~\eqref{IASD0}, both the non-perturbative dynamics on the D7-branes and the non-zero cosmological constant can be seen as sourcing imaginary anti-self-dual components of $G_3$, namely  contributions that are (1,2) and (3,0) in terms of the original almost complex structure. Away from the smeared limit considered in the previous sections, these additional components do not vanish. Furthermore, some of them  diverge when approaching $\Sigma_4$. Moreover, one also obtains new localized contributions to the ISD component of type (0,3).   

Although we do not construct the localized solution in this paper, in the following section we evaluate the action on-shell in such configuration, in order to compute the effective four-dimensional potential. In particular, we focus on the possible divergencies arising from the corresponding terms involving the new flux and pure spinor components discussed above. 


\subsection{Cancellation of divergences}

Given the presence of various terms that diverge at the location of the four-cycle $\Sigma_4$, one might worry that evaluating the full ten-dimensional supergravity action (including the D7-brane action) on the actual solution may lead to a divergent result. This issue was raised recently in \cite{Hamada:2018qef,Kallosh:2019oxv,Hamada:2019ack,Gautason:2019jwq,Hamada:2021ryq,Kachru:2019dvo}, without reaching a common conclusion about whether certain counterterms must be included or not. We now show that this question can be settled even without knowing the details of the localized solution. We will only assume that such solution exists, and that it satisfies the supersymmetry conditions given in Eqs.~\eqref{susymod}.  

Let us first present the issue at hand more explicitly. Consider, for instance, the  localized contribution to the (0,3) component of the $G_3$ flux obtained in \cite{Dymarsky:2010mf}. In terms of the first order approximation considered there, we can write $G_3^{(0,3)} \sim \svev \delta^{(0)} [\Sigma_4] \bar{\Omega}$.  Upon evaluating the flux kinetic term in the supergravity action on-shell, one picks up a contribution to the effective four dimensional potential of the form 
\begin{equation}
    \int_{M_6} G_3 \wedge \star_6 \bar{G}_3 \sim  
    \int_{M_6} |S|^2 \left( \delta^{(0)} [\Sigma_4]\right)^2 
    {\rm vol}_6
    +\cdots , 
\end{equation}
which is clearly divergent, and is furthermore difficult to interpret. Of course, this is not the only divergent term, and moreover, this is not the only type of divergence we can have: although the (1,2) component of $G_3$ given in \cite{Dymarsky:2010mf} is not localized, it still diverges at $\Sigma_4$. 
It was argued recently in \cite{Hamada:2021ryq} that the different divergent terms coming from the fluxes do not cancel out. This problem must be resolved if we expect to have a consistent picture. 
The authors of \cite{Hamada:2018qef,Hamada:2019ack} have argued that this should be understood in terms of an admixture between the conjectured "perfect square" structure of the higher-order D7-brane action \cite{Kallosh:2019oxv} and a \textit{local} renormalization procedure put forward in \cite{Hamada:2021ryq}. This is in contradiction with the analysis provided in the appendix C of \cite{Kachru:2019dvo}, where it was argued that no counterterms were necessary. 

We resolve this conundrum  by computing the on-shell action. For that, we use the expression for the four-dimensional action  in generic GCG compactifications derived in \cite{Lust:2008zd}. For the ten-dimensional metric ansatz \eqref{10Dmetric} the effective  four-dimensional action is 
\begin{equation}
\label{4d eff}
    S_{\text{eff}}=\int_{X_4}\, d^4x \sqrt{-g}\left(\frac{1}{2}\mathcal{N} R_4 -2\pi V_{\text{eff}} \right) \,,
\end{equation}
where $R_4$ is the four-dimensional scalar curvature, and the effective potential is given by the following expression
\begin{align}
\label{Veff}
    V_{\text{eff}}=&-\frac{1}{2}\int_{M_6} \, {\rm vol}_6 e^{4A} [\tilde{F}-e^{-4A}d_H (e^{4A-\Phi} \re \Psi_+) ]^2 \nonumber \\
    &+\frac{1}{2}\int_{M_6} \, {\rm vol}_6 [d_H (e^{2A-\Phi}\im \Psi_+)]^2 +\frac{1}{2} \int_{M_6} \, {\rm vol}_6 e^{-2A} |d_H (e^{3A-\Phi}\Psi_-)|^2 \nonumber \\
    &-\frac{1}{4}\int_{M_6}\, e^{-2A} \left(\frac{|\< \Psi_+,d_H(e^{3A-\Phi}\Psi_-)\>|^2}{{\rm vol}_6} +\frac{|\<\bar{\Psi}_+,d_H(e^{3A-\Phi}\Psi_-)\>|^2}{{\rm vol}_6}\right) \nonumber \\
    &+ \sum_{i \in loc. sources} \tau_i \int_{M_6}\, e^{4A-\Phi}\left( {\rm vol}_6 \rho_i^{\text{loc}}-\< \re \Psi_+,j_i\> \right) \nonumber  \\
    &-4\int_{M_6}\, \<e^{4A-\Phi}\re \Psi_+-C^{el},d_H F+j_{\text{tot}}\> \,.
\end{align}
Here, for a given polyform $A$ we have $\left[A\right]^2 { \rm vol}_6=A\wedge *_6A$ and $|A|^2\, { \rm vol}_6=A\wedge *_6\bar{A}$. On the other hand, for the expressions in the third line one should first compute the 6-form given by the Mukai pairing, and then square only the coefficient in front of the volume form. 
We consider supersymmetric solutions for which  the Bianchi Identities are satisfied and the relevant cycles are calibrated. Hence, the last two lines in Eq.~\eqref{Veff} vanish identically. 

Let us briefly compute this action on-shell in the absence of gaugino condensates, and verify that by using the supersymmetry conditions \eqref{susy}, one obtains the expected cosmological constant term. 
Upon using \eqref{susy3} in the first term of \eqref{Veff}, the integrand becomes 
\begin{align}
\label{1mu}
 e^{4A} [\tilde{F}-e^{-4A}d_H (e^{4A-\Phi} \re \Psi_+) ]^2   {\rm vol}_6&= -\frac{1}{2}e^{4A} \left[-3e^{-A-\phi}\re \left[ \bar{\mu} \, \Psi_- \right] \right]^2 {\rm vol}_6 \\
 &=18\, e^{2A-2\phi}|\mu|^2 {\rm vol}_6 \,,\nn
\end{align}
where we used the self-duality and normalization of the pure spinors, given respectively in Eqs. \eqref{*Psi} and \eqref{purespinorVol}.
For the second line of \eqref{Veff}, the first term vanishes due to \eqref{susy1}, whereas using \eqref{susy2} the second one gives
\begin{equation}
\label{2mu}
  \, e^{-2A} |d_H (e^{3A-\Phi}\Psi_-)|^2 \, {\rm vol}_6 =  \frac{1}{2}e^{-2A}|2i \mu e^{2A-\phi}\im \Psi_+|^2=8e^{2A-2\phi}|\mu|^2{\rm vol}_6 \,.
\end{equation}
Finally, it is not hard to see that, employing \eqref{susy1}, the third line contributes  
\begin{equation}
\label{3mu}
      -\frac{1}{4} e^{-2A} \left(\frac{|\< \Psi_+,2i e^{2A-\phi} \im \Psi_+\>|^2}{{\rm vol}_6} +\frac{|\<\bar{\Psi}_+,2i e^{2A-\phi} \im \Psi_+\>|^2}{{\rm vol}_6}\right) = -32 |\mu|^2e^{2A-2\phi}{\rm vol_6} \,.
\end{equation}
Putting everything together, we get 
\begin{equation}
\label{Veffng}
    V_{\text{eff}}=-6|\mu|^2 \int_{M_6}\, e^{2A-2\phi}{\rm vol}_6 \hspace{10pt} \Rightarrow \hspace{10pt} 2\pi V_{\text{eff}}=\Lambda\,\mathcal{N}\,
\end{equation}
as expected.

We now include the effect of the localized gaugino condensate. Besides the same three terms proportional to $|\mu|^2$, using the modified supersymmetry conditions \eqref{susymod} in the same way as above, new terms proportional to either $\delta^{(0)}$ or $(\delta^{(0)})^2$ will be generated. More explicitly, the on-shell value of the contributions to the effective potential contained in Eq.~\eqref{Veff} now gives
\begin{align}
\label{Veffwg}
V_{\text{eff}}^{\rm bulk}=&-\frac{1}{2}\int_{M_6}\, {\rm vol}_6 \left[-3e^{-A-\phi}\re (\bar{\mu}\Psi_-)+e^{-3A}\delta^{(0)}\left[ \Sigma_4\right]\re \left[\bar{\svev} \Psi_-\right] \right]^2   \nonumber \\
&+\frac{1}{2}\int_{M_6}\, {\rm vol}_6 e^{-2A} \left|2i\mu e^{2A-\phi}\im \Psi_+ - 2 i \svev \delta^{(2)}\left[ \Sigma_4\right]\right|^2 \nonumber \\
&-\frac{1}{4}\int_{M_6}\, e^{-2A} \left(\frac{|\< \Psi_+,2 i \mu e^{2A-\phi} \im \Psi_+  - 2 i \svev \delta^{(2)}\left[ \Sigma_4\right]\>|^2}{{\rm vol}_6} \right. \nonumber \\
& \hspace{75pt}+\left. \frac{|\< \bar{\Psi}_+,2 i \mu e^{2A-\phi} \im \Psi_+  - 2 i \svev \delta^{(2)}\left[ \Sigma_4\right]\>|^2}{{\rm vol}_6} \right).
\end{align}
%
%
This expression can be evaluated using the properties given in Eqs.~\eqref{purespinorVol}, \eqref{*Psi}, \eqref{compat}, together with the definitions  \eqref{delta2} and \eqref{delta0calibrated}. The contributions to the different type of terms we obtain for the integrand of \eqref{Veffwg} are summarised in Table \ref{Table1}.  
\begin{table}[h]
\begin{center}
\begin{tabular}{ |c|c|c|c|  } 
 \hline
  & $|\mu|^2 e^{2A-2\phi} {\rm vol}_6$ & $\delta^{(0)}\left[ \Sigma_4\right] \re \left[\bar{\mu} S \right] e^{-\phi}{\rm vol_6}$ & $\left( \delta^{(0)}\left[ \Sigma_4\right]\right)^2 |S|^2 e^{-2A}{\rm vol_6}$ \\ 
 \hline 
 1st & 18 & -12 & 2 \\ 
 \hline
 2nd & 8 & -4 & 2/3 \\ 
 \hline
 3rd & -32 & 16 & -2 \\
 \hline
 \textbf{Total} & \textbf{-6} & \textbf{0} & \textbf{2/3} \\
 \hline
\end{tabular}
\caption{The different contributions to the bulk on-shell action of the terms in the first three lines of \eqref{Veffwg}.}
\label{Table1}
\end{center}
\end{table}
There, ``1st'', ``2nd'', and ``3rd'' indicate the contributions from the first, second, and third lines on the RHS of Eq.~\eqref{Veffwg}, respectively, while the coefficients in each column give the contributions to each of the different types of terms. For instance, the numbers 18, 8, and -32 in the first column are precisely the original contributions obtained in Eqs.~\eqref{1mu}-\eqref{3mu}.  


Hence, we find that, in the on-shell bulk action, the terms proportional to $\delta^{(0)}\left[ \Sigma_4\right]$ cancel out. However, this does not happen with the divergent contributions, i.e.~those that come with a factor $\left( \delta^{(0)}\left[ \Sigma_4\right]\right)^2$. These are both unexpected results in some sense. The cancellation of terms with a single delta function in the bulk action is unexpected as 
one should also consider the brane action when the gaugino condensate acquires a non-trivial expectation value, which evaluated on-shell gives a localised contribution with a single delta. This should somehow be cancelled in order to get the desired result. On the other hand, we will see that, at least at quadratic level in the gauginos, the D-brane action does not contain terms with a square of delta functions, which can only be cancelled by counterterms, as put forward in \cite{Hamada:2021ryq}. We now discuss these two issues separately.

The brane action contains the gaugino mass-term, whose off-shell form was computed in \cite{Grana:2020hyu}, and is given in Eqs.~\eqref{4daction} and \eqref{gauginomass1}. We now evaluate this on-shell. As we already noted, the integrand in \eqref{gauginomass1} is proportional to that of the superpotential \eqref{W10D}. Hence, we can use directly Eq.~\eqref{WindetrandMU}. This holds even when using the modified supersymmetry conditions \eqref{susymod} instead of the original ones in Eq.~\eqref{susy} because the localized contributions cancel each other, see the discussion around \eqref{Wonshelllocterms}. The gaugino mass contribution to the effective potential is therefore 
\begin{equation}
\label{onshellgaugino}
   V_{\rm eff}^{\rm \lambda \lambda} = -\frac{1}{4\pi} \left( m_{\lambda} \bar{\lambda}_- \lambda_+ + c.c. \right) =-4 \int_{M_6} \,\delta^{(0)}\left[ \Sigma_4\right]e^{-\phi}\re \left[\mu \bar{\svev} \right] {\rm vol}_6 \,,
\end{equation}
where we used that $\bar{\lambda}_-\lambda_+=i16\pi^2 \bar{S}$.
Note that having a non-zero mass for the gaugino does not contradict the fact that the solution is supersymmetric since, in this context, the gaugino bilinear itself has acquired a non-trivial expectation value\footnote{In the limit where $\svev = 0$ we recover solutions with $\mu=0$, for which the mass term indeed vanishes, as expected. Higher order terms in the action would be necessary to compute the effective mass of the fermion fluctuations around the KKLT-AdS vacuum. As discussed in the main text, for the D7-brane these are difficult to compute. }.   

The D-brane action also contains higher-order terms such as terms quartic in the gauginos. These are, however, much more difficult to obtain in general, see for instance the recent computation in  \cite{Retolaza:2021jjw} of four-fermion terms in the M2 action. Moreover, there are possible  counterterms. Here we will be agnostic about how all these terms look like off-shell. Nevertheless, we note that consistency of the overall procedure demands that adding up all contributions to the on-shell action gives only the correct cosmological constant term, as in Eq.~\eqref{Veffng}. 

Combining our results given in Eq.~\eqref{onshellgaugino} and in Table \ref{Table1}, we conclude that all terms not included in our analysis above must provide two types of contributions. 
We find that a divergent contribution coming from the aforementioned counterterms must be included in order to cancel the terms proportional to $\left(\delta^{(0)}\left[ \Sigma_4\right]\right)^2$  in the third column of Table \ref{Table1}. 
Finally, four-fermion terms in the D7-brane action  (and possibly finite contributions coming from the counterterms) should add up to 
\begin{equation}
\label{fourfermion}
  V_{\rm eff}^{\lambda^4+{\rm c.t.}}=  \frac{\gamma}{64\pi^4}\int_{M_6}\, \delta^{(0)}\left[ \Sigma_4\right]e^{-\phi} |\lambda \lambda|^2  {\rm vol_6}  \,,
\end{equation}
which, using \eqref{muS}, cancels exactly \eqref{onshellgaugino}. We note here that this term is somewhat similar to the four-fermion term considered in \cite{Kachru:2019dvo}. 

Putting everything together, the effective potential in \eqref{4d eff} should be 
\begin{equation}
    V_{\text{eff}}^{\rm bulk} + V_{\rm eff}^{\rm \lambda \lambda}  +\frac{\gamma}{64\pi^4}\int_{M_6}\, \delta^{(0)}\left[ \Sigma_4\right]e^{-\phi}|\lambda \lambda|^2 {\rm vol_6} -\frac{2}{3}\int_{M_6}e^{-2A}|\svev|^2(\delta^{(0)}\left[ \Sigma_4 \right])^2 {\rm vol_6}\,,
    \label{VeffFinal}
\end{equation}
where $ V_{\text{eff}}^{\rm bulk}$ is given by the "Total" row in Table 1, and  $V_{\rm eff}^{\rm \lambda \lambda}$  is given in \eqref{onshellgaugino}. In particular, the numerical coefficients in \eqref{VeffFinal} are such that no "perfect square" structure arises.

\section{Conclusions}
\label{sec:conclusions}

In this paper we considered the ten-dimensional description of KKLT-AdS vacua. For this, we argued that the set of supersymmetry conditions in Eqs.~\eqref{susymod} describes ${\cal{N}}=1$ generalized complex geometry compactifications of type II superstring theories, including the effects of gaugino condensates on stacks of D-branes wrapping calibrated cycles of the internal manifold. In the type IIB setting, such non-perturbative contributions provide a mechanism for the stabilization of K\"ahler moduli, while the complex structure and axio-dilaton moduli acquire masses generated by 3-form fluxes\footnote{The expectation that fluxes can give masses to a large number of complex structure moduli has been challenged though by the so-called Tadpole conjecture \cite{Bena:2020xrh}.}.  

The gaugino condensate terms in Eqs.~\eqref{susymod} combine several ingredients discussed in the literature in one form or another. Eq.~\eqref{susy1mod} was put forward in \cite{Koerber:2007xk,Dymarsky:2010mf} and later discussed in \cite{Bena:2019mte}. It can be understood as an F-flatness condition for the (complexified) volume of the cycle wrapped by the D-branes undergoing gaugino condensation, if one takes into account the dependence of the Veneziano-Yankielowicz superpotential \eqref{WVY} on this modulus, which sets the value of the corresponding effective gauge coupling. On the other hand, the localized contribution in Eq.~\eqref{susy3mod} constitutes a generalization of the proposals of \cite{Dymarsky:2010mf,Kachru:2019dvo}. We have shown that it generates the correct additional term in the flux equations of motion arising when the gaugino bilinear on the branes has a non-trivial expectation value. The coupling comes from the gaugino mass term, whose precise form was obtained in \cite{Grana:2020hyu}. 

As established in \cite{Koerber:2007xk,Dymarsky:2010mf,Bena:2019mte}, a localized source with a gaugino condensate requires going beyond internal manifolds with SU(3) structure. Appropriate configurations with a more general structure group are, however, difficult to construct in practice. We have by-passed this issue by smearing the D7-branes along the internal directions. In this way, we focused on zero modes on the internal manifold, which provide the relevant ingredients for the low-energy effective four-dimensional theory. 

 We have provided an explicit ten-dimensional solution in the smeared approximation, where the extended directions span an AdS$_4$ space, while the internal manifold remains (conformally) CY. Moreover, the three-form flux is still imaginary self-dual, but it contains a crucial contribution of type (0,3), proportional to the cosmological constant. We have also shown that the latter is set by the expectation value of the gaugino condensate and the stabilized four-cycle volume. This precisely reproduces the results of \cite{Kachru:2003aw}. Importantly, we emphasize that given that the gaugino condensate generates  (0,3) fluxes (together with the cosmological constant) while keeping supersymmetry, one should not think of the AdS vacuum as the result of a two-step procedure, the first involving  supersymmetry-breaking fluxes in a Minkowski solution, and the second one adding the gaugino condensate. Clearing this misconception furthermore  avoids the criticism of \cite{Sethi:2017phn} regarding adding non-perturbative effects on top of a supersymmetry-breaking, rolling solution. This perspective was also advocated in \cite{Lust:2022lfc}.

We have also considered the issue of scale separation in this context. At least at the level of our smeared solution, we have found no obstruction for an exponentially small cosmological constant generated by the non-perturbative effects, while retaining a large internal volume. This holds as long as the fluxes can be combined in such a way that they result in an exponentially small (0,3) component. Explicit examples were provided recently in \cite{Demirtas:2019sip,Demirtas:2020ffz,Demirtas:2021nlu,Demirtas:2021ote}. These examples were however questioned in \cite{Lust:2022lfc}, where the authors argue that the cycles dual to the corresponding fluxes can not have a Special Lagrangian representative, and thus no dual brane domain-wall interpretation. We found, in the smeared, limit that  the (0,3) fluxes are given by \eqref{HOMsmeared}, suggesting that their dual cycles are indeed Special Lagrangian.  

Finally, we have discussed the localized solution, focusing in particular on the issue of divergencies arising in the on-shell evaluation of the ten-dimensional action that gives the effective potential of the four-dimensional theory. We have shown that
this can be evaluated \textit{without knowing the details of the localized solution}. Indeed, assuming that such localized solution exists, we have evaluated the expression for the effective potential in terms of derivatives of the pure spinors given in \cite{Lust:2008zd}, using only the supersymmetry conditions with gaugino condensates. We find that no "perfect square" structure is present, contrary to the expectation in  \cite{Kallosh:2019oxv,Hamada:2018qef,Hamada:2019ack}, based on four-dimensional supergravity, as well as on heterotic and type I actions. Furthermore, divergencies coming from squared delta functions do arise, indicating the need for a local counterterm. In this sense, our results suggest a structure similar to what was discussed recently in \cite{Hamada:2021ryq}, as opposed to the conclusions of \cite{Kachru:2019dvo}. Although we can precisely establish what the counterterm gives on-shell, its off-shell form remains an open question. It would be very interesting to figure out what kind of off-shell terms in the brane action would contribute to the on-shell expression we found.

\medskip

\noindent {\bf \large Acknowledgments\vspace{0.1in}}\\
We would like to thank 
Iosif Bena, Severin L\"ust, Liam McAllister, Ander Retolaza, Gary Shiu and specially Luca Martucci for several very useful discussions, as well as Gabriele Lo Monaco
for collaboration during initial stages of this work.  This work was partly supported by the ERC Consolidator Grant 772408-Stringlandscape.

\appendix

\bibliographystyle{JHEP}
\bibliography{refs}

\end{document}